\documentclass{article}

\newcommand{\be}{\begin{equation}}
\newcommand{\ee}{\end{equation}}

\begin{document}


\title{A simple frequency approximation formula for a class of nonlinear oscillators}

\author{K. Rapedius,\\
Max-Beckmann-Str.35, D-76227 Karlsruhe, Germany\\
e-mail:kevin.rapedius@gmx.de
}

\maketitle

\begin{abstract}
An astonishingly simple analytical frequency approximation formula for a class of nonlinear oscillators with large amplitudes is derived and applied to various example systems yielding useful quick estimates.
\end{abstract}

%


\section{Introduction}
\label{sec:intro}

In addition to established methods like Harmonic Balance, Krylov Bogoliubov or Lindsted Poincare \cite{MickensBook} many new approaches for approximating the limit cycle frequencies of strongly nonlinear oscillators have been introduced in recent years, e.g. the Energy Balance method \cite{He_EBM}, the Hamiltonian Approach \cite{He_HA}, the Variational Iteration method \cite{He_VIM}, the Amplitude frequency formulation \cite{He_AF} or the Newton Harmonic Balance Method \cite{Wu_NewtonHB} and other methods \cite{Ren,He_AF}. These new methods have been successfully applied to various systems (see e.~g.~\cite{Akbarzade,Durmaz2012,AkbarzadeKhan,Ganji,Naggar,Ghadimi,Tao}). 
Here, we present an extremely simple straightforwardly applicable analytical frequency approximation formula which yields satisfactory results for a variety of systems and parameter ranges with a minimum of effort. It is particularly useful in cases where the potential (or antiderivative of the force function) cannot be written in an analytically closed form.

As a first example, let us consider the Duffing oscillator 
\be
   \ddot{x}+\alpha x+\epsilon x^3=0 
   \label{DuffingDGL}
\ee 
with the initial conditions $x(t=0)=A$, $\dot{x}(t=0)=0$. The ansatz 
\be 
  x(t)=A \cos(\omega t)
  \label{Acos}
\ee 
satisfies the initial conditions and becomes an exact solution in the linear case $\epsilon=0$ if $\omega^2 =\alpha$.
Inserting our ansatz (\ref{Acos}) into the differential equation (\ref{DuffingDGL}) we obtain
\be
  -\omega^2 A \cos(\omega t) + \alpha A \cos(\omega t) + \epsilon A^3 \cos^3(\omega t) = 0 
   \label{DuffingDGL2}
\ee 
where the cubic cosine function can alternatively be written as $\cos^3(\omega t)=\frac{1}{4}\big(3\cos(\omega t)+\cos(3 \omega t) \big)$ \cite{Abramowitz}. 
Due to the term proportional to $\cos(3 \omega t)$ our ansatz $x=A \cos(\omega t)$ cannot be an exact solution of equation (\ref{DuffingDGL}). 
We seek an approximate solution for the frequency $\omega$ by means of a colocation method, i.e. by evaluating equation (\ref{DuffingDGL2}) at some time $t \in [0, T/4]$ where $T=2\pi/\omega$, similar to the procedure used in \cite{He_EBM} in the context of the Energy Balance method. In \cite{He_PRL} He used an analogous approach in combination with a Galerkin method rather than colocation. 

We want to choose our colocation time $t$ such that the influence of the $\cos(3 \omega t)$-term is small. We therefore evaluate the differential equation where $\cos(3 \omega t)=0$ for the first time which leads to the condition $3 \omega t=\pi/2$ or 
\be
  \omega t= \frac{\pi}{6} \,.
  \label{CollocTime}
\ee
The colocation point $\omega t=\pi/6$ was also successfully used, 
on a purely penomenological basis,
in the context of He's Amplitude Frequency formulation \cite{Geng,Naggar} (see also next section and the appendix).
Inserting condition (\ref{CollocTime}) into equation (\ref{DuffingDGL2}) we obtain
\be
  \omega=\sqrt{\alpha+\epsilon A^2 \cos^2\left(\frac{\pi}{6}\right)}=\sqrt{\alpha+\frac{3}{4}\epsilon A^2}
	\label{Duff_app}
\ee 
with $\cos(\pi/6)=\sqrt{3}/2$.
This approximate result coincides with other approaches like first order Harmonic Balance \cite{MickensBook}, first orders of He's Energy Balance method \cite{He_EBM} and his Hamiltonian approach \cite{He_HA} as well as other methods \cite{MickensBook,He_AF,Wu_NewtonHB,Ren}. 

The energy of the Duffing oscillator (\ref{DuffingDGL}) is given by $E=\frac{1}{2} {\dot x}^2+V(x)$ with the potential $V(x)=\frac{\alpha}{2}x^2+\frac{\epsilon}{4}x^4$. This leads to the expression
$\omega_{\rm exact}=2\pi/T_{\rm exact}$ where \cite{NayfehBook} $T_{\rm exact}=\int_{-A}^{A}\sqrt{\frac{2}{E-V(x)}} {\rm d}x=\frac{4}{\sqrt{\alpha+\epsilon A^2}}K(m)$, $m=\frac{\epsilon A^2}{2(\alpha+\epsilon A^2)}$ and $K(m)$ is the complete elliptic integral of the first kind.

Comparing the approximation (\ref{Duff_app}) with the exact frequency $\omega_{\rm exact}$ one can show \cite{He_EBM} that
the relative error of $\omega$ is always less than $7.6 \%$ even in the extreme large amplitude limit $\epsilon A^2\rightarrow \infty$.

The rest of the artricle is organized as folows: In section \ref{sec:formula}, generalizing the calculation from the introductory example a simple frequency approximation formula for nonlinear oscillators with antisymmetric force terms $f(x)$ which have a Taylor expansion at $x(0)=A$ is derived. In section \ref{sec:examples} the application of the formula is illustrated by means of several examples, including nonlinear oscillators described by discontinuous force functions and force functions without closed form potential energies. In section \ref{sec:asymmetric} the method is generalized to asymmetric force functions. The main results are summarized in section \ref{sec:conclusion}.
In appendix \ref{sec_app} we demonstrate that the frequency approximation formula derived in this paper is mathematically equivalent to a certain phenomenological variety of He's amplitude frequency formulation.
  
\section{Simple frequency approximation formula}
\label{sec:formula}
Now we consider a more general nonlinear oscillator of the type
\be
   \ddot{x}+f(x)=0 
   \label{GeneralDGL}
\ee 
with the initial conditions $x(t=0)=A$, $\dot{x}(t=0)=0$ and where $f(x)$ is antisymmetric in $x$, i.e. $f(-x)=-f(x)$ and differentiable in $x \ne 0$. Due to the symmetry properties of the system the Fourier expansion 
\be 
  x(t)=\sum_{k=1}^{\infty} b_{2k-1} \cos((2k-1)\omega t)
	\label{Fourier}
\ee
contains only odd multiples of $\omega t$ \cite{MickensBook} (see also the discussion in \cite{Wu_NewtonHB} where the same class of systems is considered). Thus the leading and next to leading order terms are $\cos(\omega t)$ and $\cos(3 \omega t)$ respectively.

Let us further assume that the force function $f(x)$ can be Taylor-expanded in the vicinity of the initial amplitude $x(0)=A$, $f(x)=f(A)+f'(A)(x-A)+ O\left((x-A)^2\right)$. 
The Taylor expansion of $x(t)$ can be written as $x(t)=A-\frac{1}{2}a_2 t^2+\frac{1}{4!}a_4t^4+O(t^6)$. Inserting these expansions in (\ref{GeneralDGL}) yields $a_2=f(A)$, 
$a_4=f'(A)$ leading to $x(t)=A-\frac{1}{2}f(A)t^2+\frac{1}{4!}f'(A)f(A)t^4+O(t^6)$. Comparing this expression with the Taylor expansion of the Fourier representation in (\ref{Fourier}) we obtain $A=\sum_{k=1}^\infty b_{2k-1}$, $f(A)=\omega^2 \sum_{k=1}^\infty (2k-1)^2 b_{2k-1}$ and $f'(A)f(A)=\omega^4 \sum_{k=1}^\infty (2k-1)^4 b_{2k-1}$. Convergence of the second (resp. third) of these series implies $\lim_{k\rightarrow \infty} (2k-1)^2 b_{2k-1}=0$ (resp. $\lim_{k\rightarrow \infty} (2k-1)^4 b_{2k-1}=0$) such that $b_{2k-1}\rightarrow 0$ faster than $1/(2k-1)^2$ (resp. $b_{2k-1}\rightarrow 0$ faster than $1/(2k-1)^4$) in the limit $k\rightarrow \infty$. This motivates approximations based on a truncation of the Fourier series (\ref{Fourier}) after a finite number of terms.

As in the introductory example we insert the ansatz $x(t)=A \cos(\omega t)$ into our differential equation (\ref{GeneralDGL}) arriving at
\be
-\omega^2 A \cos(\omega t)+f(A \cos(\omega t))=0 \,.
\ee
In analogy to the introductory example we colocate at $\omega t=\pi/6$, where the next to leading order terms proportional to $\cos(3\omega t)$ are zero, which leads to the simple approximation formula
\be
  \omega=\sqrt{\frac{f \left(\frac{\sqrt{3}}{2}A \right)}{\frac{\sqrt{3}}{2}A}} \, .
	\label{formula}
\ee
In appendix \ref{sec_app} we show that this approximation formula is mathematically equivalent to a certain variety of He's amplitude frequency formulation based on phenomenologically chosen parameters.

\section{Example applications}
\label{sec:examples}

\subsection{Example 1}
\label{subsec:cubic_quintic}
The cubic quintic oscillator with the force function
\be
  f(x)=\alpha x+ \epsilon x^3+ \lambda x^5
	\label{cubic_quintic}
\ee
reduces to the Duffing Oscillator (\ref{DuffingDGL}) in the limit case $\lambda=0$. The simple approximation formula (\ref{formula}) yields the frequency
\be
  \omega=\sqrt{\alpha+\epsilon \frac{3}{4}A^2+\lambda \frac{9}{16}A^4}\,.
  \label{Om_cubic_quintic}
\ee
A comparison with numerically exact frequencies for different values of $\lambda$ ( table \ref{tab:quintic}) reveals a good agreement.

\begin{table}
\begin{tabular}{llll}
$\lambda$ & $\omega_{\rm RK}$ \cite{Durmaz2012} & $\omega_{\rm approx}$ & Error $(\%)$\\
\hline
1 & 2.2798 & 2.3049 & 1.1010\\
5 & 2.7318 & 2.7500 & 0.6662\\
10 & 3.2057 & 3.2210  & 0.4773\\
100 & 7.7762 & 7.8102  & 0.4372\\
1000 & 23.7999 & 23.8170 & 0.0718\\
\end{tabular}
\caption{The approximate frequency $\omega_{\rm approx}$ from (\ref{Om_cubic_quintic}) for the cubic-quintic oscillator is compared with frequencies $\omega_{\rm RK}$ from numerically exact Runge Kutta calculations \cite{Durmaz2012} for different values of $\lambda$ and $\alpha=1$, $\epsilon=5$, $A=1$. }
\label{tab:quintic}
\end{table}

\subsection{Example 2}
\label{subsec:fractional}
The fractional strongly nonlinear oscillator described by
\be 
  f(x)=x^{1/3} \, ,
  \label{fractional}
\ee
has been considered in several articles \cite{Akbarzade,Belendez,Mickens2001,Mickens2002}. From equation (\ref{formula}) we obtain the approximate frequency
\be
  \omega=\left(\frac{4}{3}\right)^{1/6}A^{-1/3}\approx 1.0491 A^{-1/3} \, 
  \label{Om_fractional}
\ee
which coincides with the first order Harmonic Balance result \cite{Mickens2001}. A comparison with the exact frequency $\omega_{\rm ex}=1.070451 A^{-1/3}$ \cite{Belendez} reveals an error of approximately $2.0 \%$.

\subsection{Example 3}
\label{subsec:Inverse}
Next we consider the strongly nonlinear oscillator with
\be
   f(x)=x^{-1}
	\label{Inverse}
\ee
analyzed in \cite{Akbarzade,Mickens2007,He_AF}. Formula (\ref{formula}) yields the approximation
\be
  \omega=\frac{2}{\sqrt{3}}A^{-1} \approx 1.1547 A^{-1} \, 
  \label{Om_Inverse}
\ee
coinciding again with the first order Harmonic Balance result \cite{Mickens2007} and results from He's homotopy perturbation method \cite{He_AF}. The exact frequency reads 
$\omega_{\rm ex}= \frac{\sqrt{2\pi}}{2} A^{-1} \approx 1.2533141 A^{-1}$ \cite{Mickens2007}. The resulting error of $7.9\%$ is acceptable considering the simplicity of our approach.

\subsection{Example 4}
\label{subsec:MassOnWire}
An oscillator with the force function
\be 
  f(x)=x-\lambda \frac{x}{\sqrt{1+x^2}}
	\label{MassOnWire}
\ee	
was used in \cite{BelendezWire, Shou, Sun2007} to model the dynamics of a mass attached to a stretched wire. Using formula (\ref{formula}) we obtain the approximate frequency
\be
  \omega=\sqrt{1-\frac{\lambda}{\sqrt{1+(3/4) A^2}}} \, ,
  \label{Om_MassOnWire}
\ee
coinciding with the result obtained in \cite{BelendezWire} using a Harmonic Balance approach, which was shown to be in good agreement with the exact frequencies yielding relative errors no larger than $2.7\%$ for $0 \le \lambda \le 1$.

%
%
%

\subsection{Example 5}
\label{subsec: sin(x3)}
For the nonlinear oscillator with the force function
\be
   f(x)=\sin(x^3), \quad 0 \le x^3 \le pi/2
	\label{sin3}
\ee
the potential energy cannot be written in an analytically closed form. The approximation (\ref{formula}) reads
\be
  \omega=\sqrt{2\frac{\sin(3^{3/2}A^3/8)}{\sqrt{3}A } } \,.
	\label{om_sin3}
\ee
A comparison with Runge Kutta calculations (table \ref{tab:Example5}) shows the validity of the approximation (\ref{om_sin3}) for values $0\le x \le 1$.  

\begin{table}
\begin{tabular}{llll}
$A$ & $\omega_{RK}$ & $\omega_{\rm approx}$ & Error $(\%)$\\
\hline
0.1 & 0.0847 & 0.0866  & 2.2 \\
0.5 & 0.423 & 0.433 & 2.4\\
0.75& 0.630 & 0.645  & 2.5\\
1 &  0.806 &  0.836 & 3.7\\
\end{tabular} 

\caption{The approximate frequency $\omega_{\rm approx}$ from (\ref{om_sin3}) for Example 5 is compared with results of a Runge Kutta integration $\omega_{RK}$ 
for different values of $A$. 
}
\label{tab:Example5}
\end{table}

\subsection{Example 6}
\label{subsec:discont}
Next we consider the force function
\be 
  f(x)={\rm exp}(\beta \sqrt{|x|}) {\rm sign}(x) 
	\label{discont}
\ee	
which is discontinuous at $x=0$. Furthermore it potential energy cvannot be written in terms of elementary functions if $\beta \ne 0$ (where ${\rm sign}(x)= 1$ if $x > 0$, ${\rm sign}(x)=-1$ if $x \le 0$). Formula (\ref{formula}) leads to the approximation
\be
   \omega=\sqrt{ \frac{{\rm exp}\left(\beta \sqrt{\frac{\sqrt{3}}{2}A}\right)}{\frac{\sqrt{3}}{2}A} }\,.
		\label{om_discont}
\ee
For $\beta=0$, comparison with the exact solution \cite{Liu} $\omega_{exact}=\frac{\pi}{2\sqrt{2A}}$ yields an error of approximately $3.3$ \%.
Comparison with Runge Kutta calculations for $\beta=\pm 1$ (see table \ref{tab:Example6}) also reveals a decent agreement.

\begin{table}
\begin{tabular}{llll}
$\beta=-1$: & & &\\
\hline
$A$ & $\omega_{RK}$ & $\omega_{\rm approx}$ & Error $(\%)$\\
\hline
0.1 & 3.0448 &  2.9331 & 3.7\\
1   & 0.7073 &  0.6748 & 4.6\\
5   & 0.1828 &  0.1698 & 6.3\\
10  &  0.0845 & 0.0780 & 7.6\\

\end{tabular} 
\begin{tabular}{llll}
$\beta=1$: & & &\\
\hline
$A$ & $\omega_{RK}$  & $\omega_{\rm approx}$ & Error $(\%)$\\
\hline
0.1 & 4.0525 & 3.9367 &  3.0\\
1 &  1.7470 & 1.7112  & 2.1\\
5 &   1.3713 & 1.3602  & 0.81 \\
10 &  1.4813 & 1.4800 & 0.088\\
\end{tabular} 


\caption{The approximate frequency $\omega_{\rm approx}$ from (\ref{om_discont}) for Example 6 is compared with results of a Runge Kutta integration $\omega_{RK}$ 
for different values of $A$ with $\beta=-1$ and $\beta=1$. }
\label{tab:Example6}
\end{table}

\section{Generalization to asymmetric force functions}
\label{sec:asymmetric}
Now we drop the requirement of antisymmentry and consider a more general oscillator with rest position $x=0$, $f(x)$ continuous if $x\ne 0$ and Taylor-expandable about 
$x=\pm A$. According to formula (\ref{formula}) we define the frequencies
\be
  \omega_\pm=\sqrt{\frac{f \left(\frac{\sqrt{3}}{2} x_\pm \right)}{\frac{\sqrt{3}}{2}x_\pm}} \, 
	\label{formula_pm}
\ee
where $x_\pm$ denote the right and left turning points. We consider the initial conditions $x(0)=x_+$, $\dot x(0)=0$.
Following Hu \cite{Hu} we decompose the half period $T/2$ into two parts: the elapsed time between $x=x_+$ and $x=0$ is approximately given by $\frac{\pi}{2 \omega_+}$, the time between $x=0$ and x=$x_-$ by $\frac{\pi}{2 \omega_+}$. 
Thus 
\be 
  T=\frac{\pi}{\omega_+}+\frac{\pi}{\omega_-}
	\label{T_asymm}
	\ee
is an approximate value for the period and
\be 
   \omega=2 \left(\frac{1}{\omega_+}+\frac{1}{\omega_-} \right)^{-1}
\ee
for the frequency of the oscillator. 

As an example we consider the asymmetric oscillator with $f(x)=x+x^2$ with initial conditions $x(0)=x_+=A>0$, $\dot x(0)=0$. Energy conservation $E=\frac{1}{2}x_\pm^2+\frac{1}{3}x_\pm^3$ leads to \cite{Hu} 
\be
  x_-=-\left(\frac{3}{4}+\frac{A}{2} \right)+\sqrt{\left(\frac{3}{4}+\frac{A}{2} \right)^2-\left(\frac{3}{2}A+A^2\right)} \,.
	\label{Xminus}
\ee
For the existence of a periodic solution the discriminant in (\ref{Xminus}) must be positive which leads to the condition $0< A <1/2$. In table \ref{tab:asymmetric}
we compare the approximation (\ref{T_asymm}) with exact results \cite{Hu} for several values of $A$, obtaining a good agreement. 

\begin{table}
\begin{tabular}{llll}
$A$ & $T_{\rm exact}$ \cite{Hu} & $T_{\rm approx}$ & Error $(\%)$\\
\hline
0.1 & 6.3116 & 6.3122 & 9.506 e-3\\
0.2 & 6.4414 & 6.4140 & 4.254 e-1\\
0.3 & 6.6294 & 6.6354  &9.051 e-2 \\
0.4 & 7.1246 & 7.1318  &1.011 e-1 \\
0.45 & 7.7065 & 7.6967 &1.272 e-1 \\
0.49 & 9.2080 & 8.9909 &2.358 e+0 \\
\end{tabular}
\caption{The Period $T_{\rm approx}$ from (\ref{T_asymm}) with right turning point $x_+=A$ and left turning point (\ref{Xminus}) for the asymmetric oscillator with $f(x)=x+x^2$ is compared with exact results $T_{\rm exact}$ \cite{Hu} for different values of $A$. }
\label{tab:asymmetric}
\end{table}

\section{Conclusion}
\label{sec:conclusion}
A simple frequency approximation formula for a class of strongly nonlinear oscillators with antisymmetric position-dependent force terms was derived and applied to several example systems. This derivation provides a justification for a hitherto only phenomenologically applied variety of the amplitude frequency formulation which is shown to be mathematically equivalent to the formula presented here. The formula yields decent to good results for various systems with a minimum of effort. Even in the case of force functions without closed-form potential energies or discontinuous force functions approximate frequencies can be easily obtained in an analytically closed form. It was further shown that the method can be extended to asymmetric force functions.



\appendix

\section{Equivalence between the simple frequency approximation formula and a version of He's amplitude frequency formulation}
\label{sec_app}
In He's amplitude frequency formulation \cite{He_AF} an approximate frequency for a nonlinear oscillator $\ddot x+f(x)=0$, $x(0)=A$, $\dot x(0)=0$ is given by

\be 
   \omega_{AF}^2=\frac{\omega_2^2R_2(t_2)-\omega_1^2R_1(t_1)}{R_2(t_2)-R_2(t_1)}
\label{omq_AF}
\ee 
where the residuals are given by $R_{1,2}(t)=\ddot(A\cos(\omega_{1,2}t))+f(A \cos(\omega_{1,2}t)$.
He's original version used the parameters $\omega_2=1$, $\omega_1=\omega$ and $t_1=t_2=0$.

In \cite{Geng} Geng and Cai obtained improved results by choosing the same trial frequencies yet, without further justification, different times 
\be
  t_{1,2}=T_{1,2}/12
	\label{Times}
\ee	
where $T_{1,2}=2\pi/\omega{1,2}$ which is equivalent to $\omega_{1,2}t_{1,2}=\pi/6$. Inserting these values into \ref{omq_AF} yields
$\omega_{AF}^2=\frac{-\omega^2 A \cos(\frac{\pi}{6})+f(A\cos(\frac{\pi}{6})) -\omega^2(-A \cos(\frac{\pi}{6})+f(A\cos(\frac{\pi}{6}) )}{-\omega^2 A \cos(\frac{\pi}{6})+f(A\cos(\frac{\pi}{6})) - (-A \cos(\frac{\pi}{6})+f(A\cos(\frac{\pi}{6}))}=\frac{f(\frac{\sqrt{3}}{2}A)}{\frac{\sqrt{3}}{2}A}$
coinciding with (\ref{formula}). This explains the hitherto only phenomenologically observed effectiveness of the particular choice (\ref{Times}) for the times $t_{1,2}$.

\end{document}